\documentclass[12pt]{article}
\newfont{\Bbb}{msbm10 scaled 1200}     
\newcommand{\mathbb}[1]{\mbox{\Bbb #1}}

\def\lbldef#1#2{\expandafter\gdef\csname #1\endcsname {#2}}
\def\eqn#1#2{\lbldef{#1}{(\ref{#1})}%
\begin{equation} #2 \label{#1} \end{equation}}

\def\href#1#2{#2}

\newcommand{\beq}{\begin{equation}}
\newcommand{\eeq}{\end{equation}}
\newcommand{\ber}{\begin{eqnarray}}
\newcommand{\eer}{\end{eqnarray}}

\newcommand{\beqar}{\begin{eqnarray}}

\newcommand{\eeqar}{\end{eqnarray}}

\begin{document}
\baselineskip=15.5pt
\pagestyle{plain}
\setcounter{page}{1}
\begin{titlepage}

\leftline{\tt hep-th/0012218}

\vskip -.8cm

\rightline{\small{\tt CALT-68-2311}}
\rightline{\small{\tt CITUSC/00-066}} 

\begin{center}

\vskip 1.7 cm

{\LARGE {How  Noncommutative  Gauge  Theories}}

\vskip .3cm

{\LARGE{Couple  to   Gravity}}  

\vskip 1.5cm
{\large Yuji Okawa \ and \ Hirosi Ooguri}

\vskip 1.2cm

{California Institute of Technology 452-48, Pasadena, CA 91125}

{and}

{CIT-USC Center for Theoretical Physics, Los Angeles, CA 90089}

\bigskip
{\tt okawa, ooguri@theory.caltech.edu}

\vskip 1.7cm

{\bf Abstract}
\end{center}

\noindent
We study coupling of  noncommutative gauge theories on 
branes to closed string in the bulk. We derive an 
expression for the gauge theory operator dual to the 
bulk graviton, both in bosonic string theory and superstring
theory. In either case, we find that the coupling is 
different from what was expected in the literature 
when the graviton is polarized in the noncommutative 
directions. In the case of superstring, the expression
for the energy-momentum tensor is consistent with the way
the bulk metric appears in the action for the noncommutative 
gauge theory. We also clarify some aspects of the 
correspondence between operators in the 
gauge theory and boundary conditions in the dual 
gravitational description.

\end{titlepage}

\newpage


\section{Introduction}

Noncommutative gauge theories can be realized as  
the zero slope limit of open string theory on branes 
in a background of strong NS-NS two-form
field $B_{ij}$ [1 - 11].
It was pointed out in  \cite{Das:2000md,Gross:2000ba}  
that, for each local observable 
in a gauge theory in commutative space, one can define an
observable in the noncommutative version of the theory.
The construction is done in momentum space
and involves an open Wilson line \cite{Ishibashi:1999hs}
whose end-points
are separated by a vector $l^i$ defined by 
\eqn{vectorl}{l^i = k_j \theta^{ji},}
where $k_i$ is the momentum carried by the operator, and
$\theta^{ij}$ is the noncommutative parameter,
\eqn{thetaparameter}{ [ x^i, x^j ] = -i\theta^{ij}.}
The condition (\ref{vectorl}) is necessary in order
to maintain gauge invariance. 
Various aspects of the open Wilson line were discussed
in [15 - 19]. Open Wilson lines make an over-complete set of 
gauge invariant observables. An open 
Wilson line of any shape is gauge invariant as long as its 
end-points are separated by $l^i$. We can also insert 
operators in various ways along the line. 

In fact, there is a canonical prescription 
to define gauge invariant observables in noncommutative theories
corresponding to a large class of operators in commutative theories. 
These are operators which describe the coupling 
of the gauge theory on the brane to closed string states in the bulk. 
In this paper, we find precise expressions
of these operators in noncommutative theories
by studying disk amplitudes of one closed 
string and arbitrary number of open strings and by taking
the zero slope limit. 
Disk amplitudes with a closed string state
in the presence of a constant $B$ field were discussed 
in \cite{Hyun:1999ma,Garousi:1999ch},
and it was found that
the amplitudes involve the so-called generalized star 
products.\footnote{A similar structure also appear in 
one-loop amplitudes
of these theories 
[22 - 24].}
In \cite{Mehen:2000vs}, it was shown that these products
also appear if one considers
an open Wilson line which connects $x$ and $x+l$ 
by a straight line and expands it in powers of gauge fields.  
Combining these observations,
it was anticipated that the open Wilson lines play
some role in constructing gauge theory operators coupled 
to closed string states. The link between
generalized star products and 
open Wilson lines has been investigated 
further in [26 - 30].

In this paper, we set up our computation so that we can
determine the gauge theory operators that couple 
to arbitrary excited closed string states. We show that 
a basic ingredient of such operators is a {\it straight} open
Wilson line of the form, 
\eqn{preciseline}{
\int dx *\left[e^{ikx} \exp\left(i \int_0^1 d\tau
~l^i A_i(x+l \tau) 
+ y_\alpha \Phi^\alpha(x+l\tau) \right) \right], }
where $x$ are the coordinates on the brane, 
$A_i$ is the gauge field in the noncommutative
directions, $\Phi^\alpha$ is the scalar
field describing the transverse positions of the
branes, and $*[\cdots]$ means that we take the standard star
product in the expansion of the expression in $[\cdots]$
in powers of $A_i$ and $\Phi^\alpha$. 
The path-ordered exponential is implicit in 
this expression and throughout the rest of the paper. 
The vector $l^i$ is defined by (\ref{vectorl})
and $y_\alpha$ is given by
\eqn{vectory}
{y_\alpha = 2\pi \alpha' k_\alpha,}
where $k_\alpha$ is the momentum
in the directions transverse to the brane.
Because of the closed string on-shell condition,
the Wilson line can couple to a closed string only 
when $l^i$ and $y_\alpha$ satisfy the constraint
\eqn{constraintfirst}{
   G_{ij} l^i l^j + g^{\alpha\beta}
 y_\alpha y_\beta = 0,}
where $G_{ij}$ is the open string metric
in the noncommutative directions
and $g_{\alpha\beta}$ is the closed string metric
in the directions transverse to the brane. 

The condition (\ref{constraintfirst}) is 
independent of the amount of closed string
excitations. For coupling with excited states, 
we need to insert additional gauge theory operators 
along the Wilson line (\ref{preciseline}). We 
describe a systematic procedure to determine
these additional operators and the way they are
inserted along the Wilson line. 

We derive an explicit expression for the operator
coupled to the bulk graviton, $i.e.$, the energy-momentum
tensor. In the case of
bosonic string, the structure of the operator suggests that 
the leading coupling between the bulk metric and the 
gauge fields is through the curvature of the metric.
In the case of superstring, the leading coupling to 
the curvature is absent. 
We then go to the next leading order in the $\alpha'$ 
expansion of the disk amplitude and derive
the energy-momentum tensor. We find that the resulting
expression agrees, in the limit
of zero momentum, with the variation of
the action of the noncommutative gauge theory [10]
with respect to the bulk metric. In both bosonic
and supersymmetic cases, we find that the coupling 
between the gauge fields and the bulk metric 
is different from what was expected in the 
literature when the graviton is 
polarized in the noncommutative directions. 

A general structure of the gauge theory operators
fits well with the proposed gravity duals of the
noncommutative theories at large $N$ 
\cite{Hashimoto:1999ut,Maldacena:1999mh}. 
In a gauge theory in commutative space,
it is natural to classify 
operators into irreducible
representations of global symmetry of the
theory. For example, the four-dimensional 
${\cal N}$=4 theory has the $SO(6)$ $R$ 
symmetry, which we identify as the rotational 
group in the transverse directions to the brane. 
We classify the operators according to
the $R$ symmetry since operators with
different $R$ charges are renormalized
differently and carry different conformal weights. 
The situation changes
in noncommutative theories discussed in this paper. 
We note that the open Wilson line (\ref{preciseline})
does not belong to a definite representation 
of the $R$ symmetry, but rather it is parametrized 
by the vector $y_\alpha$ in the transverse directions. 
By expanding in powers of $y^\alpha$, one finds  
that the Wilson line is a superposition of
infinitely many representations of the $R$ symmetry. 
We show that it in fact matches well with an asymptotic 
behavior of a solution to a wave equation in the background
geometry for the dual gravitational description. 
This observation
also helps us clarify some aspects of the correspondence between 
operators in the gauge theory and boundary conditions in the
dual gravitational description. 

This paper is organized as follows.
In Section 2, we compute the disk amplitudes and derive
the coupling of the noncommutative gauge fields to closed
string states in the bulk. After a description of our set-up,
we start with a computation of a bosonic string $S$-matrix 
involving a closed string tachyon and arbitrary number
of gauge fields. By taking the zero slope limit, we 
find that the tachyon in the bulk couples to the 
straight open Wilson line (\ref{preciseline}) satisfying
the constraint (\ref{constraintfirst}). We then 
study the coupling to the graviton in the bulk
and derive the energy-momentum tensor on the brane,
in both bosonic string theory and superstring theory. 
The structure of gauge theory operators coupled to 
massive string modes is briefly described.
In Section 3, we discuss the dual 
gravitational descriptions of the noncommutative 
theories. A detailed derivation of the energy-momentum 
tensor and a proof of its conservation in the case of
bosonic string are presented in Appendices A and B.

\bigskip

After completion of the work, we were informed
that Thomas Mehen and Mark Wise have independently computed  
a disk amplitude of one graviton and two gauge fields
in bosonic string theory.
Their work overlaps with some of the results presented 
in this paper.  

\section{Disk Amplitudes and Open Wilson Lines}

In this section, we study the disk amplitude.
The worldsheet disk is represented by
the upper half-plane of the complex plane,
and the propagator
is given by [33 - 35]
\ber \label{prop}
\langle X^M(z) X^N(w) \rangle 
& = &- \alpha'\left[  g^{MN} \log|z-w| - g^{MN} \log|z-\bar{w}| +
\right.  \nonumber \\
&&~~~~~\left. +2 G^{MN} \log|z-\bar{w}|+ {1\over 2\pi\alpha'}
\theta^{MN} \log\left({z-\bar{w} \over \bar{z}-w}\right)\right].
\eer
We decompose the coordinates $X^M$ into
three directions: $M = (\alpha,\mu,i)$ where
$\alpha$ runs in the directions transverse to
the brane, 
$\mu$ and $i$ in the commutative and noncommutative directions,
respectively, along
the branes. Therefore, $\theta^{MN}=0$ unless $(M,N)=(i,j)$.
The closed string metric is denoted by $g_{MN}$,
so the open string metric $G_{MN}$ on the brane $M=\mu, i$
is given by 
\eqn{openmetric}{
  G^{\mu\nu} = g^{\mu\nu}~,~~~
G^{ij}= {1\over (2\pi \alpha')^2} \theta^{im} \theta^{jn} g_{nm}, }
in the zero slope limit \cite{Seiberg:1999vs}. We also set 
\eqn{openmetrictransverse}{
G^{\alpha\beta} =0,}
so that the propagator (\ref{prop}) obeys the correct
Dirichlet boundary conditions in the transverse directions. 
The noncommutative gauge theory is defined by taking
the zero slope limit $\alpha' \rightarrow 0$, 
while keeping $G^{ij}$ and $\theta^{ij}$ finite. 

The amplitudes we consider contain one closed string
with momentum $k$ and an arbitrary number of open strings 
with momenta $p_a$ ($a=1,\ldots,n$).
In string theory, we can only compute on-shell physical 
amplitudes. In the zero slope limit, the gauge field
$A_M$ on the brane can have any amount of momentum since 
the on-shell condition,
\eqn{openonshell}{\alpha' G^{MN}p_M p_N=0,} 
holds for any finite $p$. Another way to say this is that
the open string operator $e^{ipX}$ 
does not receive anomalous dimensions when $\alpha'=0$. 

Once we choose the open string momenta, 
the closed string momentum $k$
along the brane is determined by the momentum conservation 
$k + p_1 + \cdots + p_n=0$. Since
the inverse $g^{ij}$ of the closed string metric
scales as $\alpha'^{-2}$ in the noncommutative direction,
if $k_i$ is non-zero and finite,
$g^{ij}k_ik_j$ is large $\sim \alpha'^{-2}$
and positive. Therefore,  
to maintain the closed string on-shell condition 
\eqn{closeonshell}{\alpha' g^{MN} k_M k_N = 
4(1-{\bf N})~~~({\bf N}: 
{\rm amount~of~closed~string~excitations}),}
the closed string has to have an energy which grows as $\alpha'^{-1}$.
This is a well-known fact and reflects the decoupling of
closed string modes from the gauge
theory degrees of freedom on the brane [36 - 38].
In order to keep the momentum $k = -(p_1+\cdots+p_n)$
along the brane to be finite, 
we assume that the brane is oriented in spacelike
directions so that the $O(\alpha'^{-1})$ component of $k$
is transverse to the brane.\footnote{Alternatively
we may take the brane to be oriented in space-time
directions by allowing the transverse momentum to
be pure imaginary. This set-up is more appropriate when
we apply our results to the dual gravitational description of 
large $N$ theories \cite{Hashimoto:1999ut,Maldacena:1999mh},
which we will discuss in Section 3. 
In the dual description, the gauge theory operators are located at the 
boundary of the bulk geometry. They are in fixed 
momentum states in the noncommutative directions
but can be localized in the commutative directions.
The insertion of a gauge theory operator is realized by imposing 
a boundary condition on the corresponding bulk  
field that it grows exponentially toward the 
boundary points where the operators are inserted
but decays exponentially toward elsewhere. 
Namely, the bulk field carries an imaginary momentum 
in the transverse directions. 
Everything we say in the following is unchanged in
this set-up, except that we need to make the substitution 
$y_\alpha \rightarrow i y_\alpha$ in some of the equations. 
We may also consider the case when 
the bulk metric is Euclidean, again by allowing
the transverse momentum to be imaginary.} 
The scaling $k_\alpha \sim \alpha'^{-1}$ fits well with  
the fact that the scalar fields $\Phi^\alpha$
describing the location of the branes in the transverse
directions couple to $y_\alpha = 2\pi \alpha' k_\alpha$. 

In the limit $\alpha' \rightarrow 0$,
the on-shell condition (\ref{closeonshell}) reduces to 
\eqn{reducedonshell}
{ G_{ij} \theta^{im} \theta^{jn} k_{m} k_{n} 
+ g^{\alpha\beta} y_\alpha y_\beta = 0.}
We note that this condition is independent
of the amount ${\bf N}$ of closed string excitations.  
We can then study the coupling
of arbitrary excitations of closed string to the gauge 
theory on the brane while keeping
open string momenta to be finite and arbitrary.

It is also important to note that the momentum $k_\mu$
in the commutative directions on the brane does not appear
in the on-shell condition (\ref{reducedonshell}). 
Thus one can perform the inverse Fourier 
transformation on $k_\mu$ to obtain gauge theory operators localized
in the commutative directions. On the other hand, it
is not possible to perform the inverse Fourier transform
on $k_i$ in the noncommutative directions if one tries
to keep $y_\alpha$ fixed since $k_i$ and $y_\alpha$
are related by (\ref{reducedonshell}). 

\subsection{Tachyon in Bosonic String} 

Let us start with the one-point function of
the close string tachyon $V(z)=e^{ikX(z)}$
with arbitrary number of open strings in
bosonic string theory. 
As a warm-up exercise, we first examine the contraction of
the closed string vertex $V(z)$
with the $e^{ipX(t)}$ part of the open string
vertex, where $z$ is in the upper half plane and $t$
is on the real axis. 
In the zero slope limit,
the bulk-boundary propagator in the worldsheet
is simplified as
\eqn{openprop}{ \langle X^M(z) X^N (t) \rangle
 = -i \theta^{MN} \tau(t,z) + O(\alpha'),}
where 
the function $\tau(t,z)$ is given by   
\eqn{proceed}{\tau(t,z) = {1 \over 2\pi i} \log
\left({t-z \over t-\bar{z}}\right).}
For a fixed value of $z$, $\tau(t,z)$
 is a monotonically increasing function of $t$
and 
\eqn{jumpintau}
{\tau(\infty,z) - \tau(-\infty,z)=1.}
We choose the branch of the logarithm in (\ref{proceed})
so that $\tau(-\infty,z)=0$. 
The amplitude in the zero slope limit is then
\ber \label{commdirection}
&&(z-\bar{z})^2\langle e^{ikX(z)} e^{ip_1X(t_1)} \cdots e^{ip_nX(t_n)} 
\rangle \nonumber \\
& =& \exp\left[-{i\over 2} 
\sum_{a<b} p_{a}\theta p_{b}\epsilon(t_a-t_b)
+ \sum_a i  k \theta p_a \tau(t_a,z)\right] \delta(k+p_1+\cdots+p_n).
\eer
Here $\epsilon(t)$ is the step function which is equal
to $1$ or $-1$ for positive and negative $t$. 
The factor $(z-\bar{z})^2$ on the left-hand side
of the equation cancels a factor due to
self-contraction of the closed string
vertex operator $e^{ikX}$. We used  
the on-shell condition for the tachyon, 
\eqn{tachyonshell}{\alpha' g^{MN} k_Mk_N = 4,}
but the open string momenta $p_a$ are arbitrary. 
In fact, the right-hand side of (\ref{commdirection})
is $SL(2,R)$ invariant for any $p_a$ since
\ber
&& \sum_a k \theta p_a \tau\left({\alpha t_a + \beta\over 
\gamma t_a + \delta},
 {\alpha z+\beta \over \gamma z + \delta}\right) \nonumber \\
&=& \sum_a  k \theta p_a \tau(t_a,z) + 
 k \theta \left(\sum_a p_a \right)  {1\over 2\pi i} \log
\left( {\gamma \bar{z} + \delta \over \gamma z+\delta} 
\right) \nonumber\\
&=& \sum_a k \theta p_a  \tau(t_a,z) 
\eer
where 
\eqn{sl2r}{\alpha\delta-\beta\gamma=1,~~~~\alpha,\beta,\gamma,\delta
:~ {\rm real},}
and we used $k \theta \sum_a p_a = - k_i \theta^{ij} k_j = 0$. 
Since the open string momenta $p_a$ are arbitrary
in the zero slope limit,
we can take the inverse Fourier transform of (\ref{commdirection})
with respect to $p_a$ with the total momentum fixed and obtain, 
\eqn{coord}{
(z-\bar{z})^2
\langle e^{ikX(z)} \prod_a f_a(X(t_a)) \rangle
 = \int dx ~* \left[ e^{ikx} \prod_a f_a(x+l\tau(t_a,z))
\right]. }
Here $l^i = k_j \theta^{ji}$ and
the integral $\int dx$ is over the brane.
The symbol $*[\cdots]$ means that we evaluate 
the expression in $[\cdots]$ using the standard star 
product with the ordering along the boundary of the worldsheet
as,
\begin{eqnarray}
&& * \left[ f_1(x+\tau(t_1,z)) f_2(x+\tau(t_2,z)) \right] \nonumber \\
& =& \left. \exp\left({i \over 2}\epsilon(t_1-t_2) \theta^{ij}
 {\partial^2 \over \partial \xi^i \partial \zeta^j}\right)
f_1(x+l\tau(t_1,z) + \xi) f_2(x+l\tau(t_2,z) + \zeta)
\right|_{\xi,\zeta=0}. \nonumber \\
&& \label{whatstar}
\end{eqnarray}
We note that the locations of the open strings 
are displaced from the closed string location $x^i$
by the amount $l^i \tau(t_a,z)$, which runs from 
$0$ to $l^i$ as $t_a$ goes from 
$-\infty$ to $\infty$. 

Now we are ready to consider the vertex operator
for gauge field,  
\eqn{gauge}{U(t) = \left(
u_i(p) {dX^i(t)\over dt}+u_\mu(p) {dX^\mu(t)\over dt}
  \right)e^{ipX(t)}.}
Here $i$ and $\mu$ run in the
noncommutative and commutative directions, respectively,
along the brane. For the polarization
$u_i$ in the noncommutative directions,
we can contract
$d X^i(t)/dt$ with $e^{ikX(z)}$ using
(\ref{openprop}). 
The zero slope limit of such an amplitude is
\ber
&& (z-\bar{z})^2
\langle e^{ikX(z)} \prod_{a=1}^n e^{ip_aX(t_a)} u_{i}(p_a){dX^{i}\over dt_a}
   \rangle\nonumber \\
&=& 
\exp\left[-{i\over 2} 
\sum_{a<b} p_{a}\theta p_{b}\epsilon(t_a-t_b)
+ \sum_a ik \theta p_a  \tau(t_a,z)\right] \times \nonumber \\
&&~~~~~~~\times \prod_a ~l^i u_i(p_a)  {\partial \tau(t_a,z) \over \partial t_a}
 \delta(k+p_1+\cdots+p_n).
\eer
In the coordinate basis, this takes the form,
\ber \label{tachyongauge}
&& (z-\bar{z})^2\langle e^{ikX(z)} \prod_a
A_{i}(X(t_a)) {d X^{i} \over dt_a}
\rangle\nonumber \\ &=&
\int dx ~*\left[e^{ikx} \prod_a l^i A_{i}(x+l \tau(t_a,z))
           {\partial \tau(t_a,z)\over \partial t_a} 
\right].
\eer
We can exponentiate this result to write\footnote{
In this paper we use the point-splitting
regularization on the worldsheet, which is known to lead to the
standard description of the noncommutative gauge theory 
\cite{Seiberg:1999vs}. }
\ber \label{firstopen}&&
 (z-\bar{z})^2
\langle e^{ikX(z)} \exp\left(i \int_{-\infty}^{\infty} dt A_i(X(t)) 
{dX^i\over dt}\right)\rangle\nonumber\\ & =&
\int dx *\left[ e^{ikx} \exp \left(i \int_0^1d\tau ~l^i
 A_i(x+l \tau)   \right) 
\right].
\eer
On the right-hand side of the equation,  
we changed integration variables from $t$ to $\tau = \tau(t,z)$. 
Thus we find that the disk amplitude with a closed string
naturally leads to the {\it straight} open Wilson line
connecting $x^i$ and $x^i+l^i$. 
Following the definition of the star product in (\ref{whatstar}),
the path-ordering in (\ref{firstopen})
is such that $A_i(x)$ is to the left. 
It is clear that, when there are several branes, the 
computation including the Chan-Paton factor gives 
the contraction of the gauge group indices of $A_i$
according to this path-ordering. 

On the other hand, the polarization $u_\mu$ 
in the commutative direction costs $\alpha'$
every time we contract $dX^\mu/dt$ with the closed
string vertex $e^{ikX}$. Thus terms involving
$u_\mu$ are subleading in the $\alpha'$ expansion.

We can also consider the vertex operator for the scalar
fields 
\eqn{scalar}{ U(t) = \phi^\alpha(p) 
~ig_{\alpha\beta} \partial_\perp X^\beta  e^{ipX},}
where 
$\partial_\perp$ is the derivative 
in the direction transverse to the boundary of the
worldsheet. As in the case of the polarization in the
commutative direction along the brane, the contraction
of $ig_{\alpha\beta} \partial_\perp X^\beta$ with the
closed string vertex $e^{ikX}$ costs $\alpha'$.
However this is compensated by the fact that
the closed string momentum $k_\alpha$ in
the transverse direction scales as $\alpha'^{-1}$.
Thus disk amplitudes with
the scalar fields $\Phi^\alpha$ are of the same order 
in $\alpha'$ as those 
with the gauge field $A_i$ in the noncommutative directions. 
By computation similar to the one that led to (\ref{firstopen}),
using
\eqn{transversederivative}
{ i\partial_\perp \log | z-t_1|^2 
=  {z - \bar{z} \over (z-t_1)(\bar{z}-t_1)},}
we find
\ber \label{secondopen}&&
 (z-\bar{z})^2
\langle e^{ikX(z)} \exp\left(i
 \int_{-\infty}^{\infty}dt  ~\Phi^\alpha(X(t))~ 
i g_{\alpha\beta} \partial_\perp X^\beta
 \right)\rangle  \nonumber\\
&=&
\int dx *\left[e^{ikx} \exp \left( i\int_0^1d\tau
 ~y_\alpha  \Phi^\alpha(x+l \tau)  \right)  
\right],
\eer
where
\eqn{whaty}{y_\alpha = 2 \pi \alpha' k_\alpha,}
and $\Phi^\alpha(x)$ is the Fourier transform
of $\phi^\alpha(p)$. Again the path-ordered
exponential is implicit in this expression. 

We can combine the amplitudes for the gauge
field $(A_\mu, A_i)$ and the scalar field
$\Phi^\alpha$ to write   
\ber \label{tachyon}
&& (z-\bar{z})^2
\langle e^{ikX(z)} \exp\left(i \int_{-\infty}^{\infty}dt~
A_i(X){dX^i\over dt}+ A_\mu(X) {dX^\mu \over dt}+ \Phi^\alpha(X) 
~ig_{\alpha\beta} \partial_\perp X^\beta\right)\rangle \nonumber\\
&=&
\int dx *\left[e^{ikx} \exp \left( i\int_0^1 d\tau ~ l^i
A_i(x+l\tau)  + y_\alpha \Phi^\alpha(x+l \tau) \right) 
\right].
\eer
It is interesting to note that,
because of the on-shell condition (\ref{tachyonshell}),
$l^i =k_j\theta^{ji}$ and $y_\alpha$ in
(\ref{tachyon}) are related as
\eqn{bps}{
 G_{ij} l^i l^j + g^{\alpha \beta} 
y_\alpha y_\beta =0.}

It is straightforward to turn (\ref{tachyon})
into the string theory $S$-matrix computation
and to read off the coupling between
the closed string tachyon and the noncommutative
gauge field. Let us look at the tachyon-$A_i$
amplitude (\ref{tachyongauge}). 
Because of the $SL(2,R)$ invariance, we can
fix the locations $z$ of the closed string 
vertex and of one of the open string
vertices, say $t_1$. The ghost factor is
\eqn{ghostfactor}
{ \langle c(z) \bar{c}(\bar{z}) c(t_1) \rangle
   = (z-\bar{z})(z-t_1)(\bar{z}-t_1).}
Since
\eqn{extrafactor}
{ {\partial \tau(t_1, z) \over \partial t_1}
 = {1 \over 2\pi i } {z-\bar{z} \over (z-t_1)(\bar{z}-t_1)},}
we can combine (\ref{tachyongauge})
and (\ref{ghostfactor}) to write
\ber \label{tachyonsmatrix} &&
2\pi i\langle c(z) \bar{c}(\bar{z}) c(t_1) \rangle
\langle e^{ikX(z)}\prod_a A_{i}(X(t_a)) {d X^{i} \over dt_a}
\rangle\nonumber \\
&=& \int dx ~*\left[e^{ikx}
l^i A_i(x+l\tau(t_1,z)) 
\prod_{a=2}^n  l^j A_{j}(x+l \tau(t_a,z))
  {\partial \tau(t_a,z) \over \partial t_a}\right].
\eer
The $S$-matrix is obtained by integrating $t_2, \ldots, t_n$
along the real axis, taking into account the ordering
of the open string vertex operators. By the $SL(2,R)$
invariance, the result of the integral is independent
of $z$ and $t_1$,
\ber \label{finalsmatrix} &&
\int_{-\infty}^\infty dt_2 \cdots dt_n
~2\pi i\langle c(z) \bar{c}(\bar{z}) c(t_1) \rangle
\langle e^{ikX(z)}\prod_a 
A_{i}(X(t_a)) {d X^{i} \over dt_a}
\rangle\nonumber \\
&=& \int dx ~*\left[e^{ikx}
\prod_{a=1}^n \int_0^1 d\tau_a ~l^j A_{j}(x+l\tau_a)\right].
\eer
We can then exponentiate the result
as in (\ref{firstopen}).
The computation  involving the scalar
field $\Phi^\alpha$ is essentially the same
except that we need to use (\ref{transversederivative}) 
in place of (\ref{extrafactor}).   

The last step is to remove one-particle reducible (1PR)
parts of gauge fields from the $S$-matrix. In fact,
the 1PR parts have already been removed in our computation
at the zero slope limit. This is more or less obvious from
the fact that there is no operator product singularities
(\ref{tachyonsmatrix}) at $t_a = t_b$. To see 
this more explicitly, we need to take $\alpha'$ to be small
but non-zero. A massless pole in the $S$-matrix
arises from an integral of the form,
\eqn{masslesspole}{
 \int dt_a ~{\alpha' \over t_a-t_b} (t_a-t_b)^{2\alpha' p_a \cdot p_b} 
\sim {1 \over p_a\cdot p_b},}
where $p_a \cdot p_b = G^{ij} p_{ai} p_{bj} + G^{\mu\nu} p_{a\mu}
p_{b\nu}$. 
The factor $(t_a-t_b)^{2 \alpha' p_a \cdot p_b}$ is due to the
contraction of $e^{ip_a X}$ and $e^{ip_bX}$ at finite $\alpha'$. 
The additional factor $\alpha'/(t_a-t_b)$ comes from 
a contraction of $dX/dt$ in one of the open string vertex 
operators with another. 
This additional singularity $(t_a-t_b)^{-1}$ is needed in order
to get the massless pole. Because of the form of the propagator 
(\ref{prop}), the singularity $(t_a-t_b)^{-1}$ is always
accompanied by a factor of $\alpha'$.
In our computation, we have taken the limit $\alpha' \rightarrow 0$
before we do any integral over $t_a$'s. If we take the
limit in this order, the left-hand side of (\ref{masslesspole})
vanishes. Thus there is no 1PR parts of gauge fields in
the $S$-matrix we computed in (\ref{finalsmatrix}). 

To summarize, we have shown that the closed string tachyon 
couples to the open Wilson line defined by 
\eqn{tachyonoperator}{
{\cal O}_{tachyon} =
\int dx *\left[ e^{i(k_\mu x^\mu + k_i x^i)} 
\exp \left( i\int_0^1 d\tau ~ l^i
A_i(x+l\tau)  + y_\alpha \Phi^\alpha(x+l \tau) \right) 
\right].}
This operator is gauge invariant provided
$l^i = k_j\theta^{ji}$ \cite{Ishibashi:1999hs}. 
The vectors $l^i$ and $y_\alpha$ have to obey
the constraint (\ref{bps}) in order for the operator to
couple to the tachyon.

\subsection{Graviton in Bosonic String}

The vertex operators of massless closed string modes are
of the form, 
\begin{equation}
V(z) = h_{MN} (k) \partial X^M \bar{\partial} X^N
e^{ikX (z)}.
\end{equation}
We will consider the case when $h_{MN}$ is a symmetric
tensor, corresponding to the graviton and the dilaton 
in the bulk. We are particularly interested in the graviton
polarized in the noncommutative directions since the
corresponding gauge theory operator can be regarded
as the energy-momentum tensor $T^{ij}$ of the gauge theory. 

The Wick contraction of $\langle V(z) \times {\rm open~strings}\rangle$
generates various terms. Let us first examine the
terms that are generated by using the first two terms
in the propagator (\ref{prop}), $i.e.$,
\eqn{firsttwo} {
\langle X^M(z) X^N(w) \rangle_{the~first~two} 
 = - \alpha'  g^{MN} \log\left| {z-w \over z-\bar{w}} \right|.}
This vanishes when either $z$ or $w$ is on the boundary,
so we can use this only for the self-contraction of
the closed string vertex operator $\partial X^M \bar{\partial} X^N
e^{ikX (z)}$. After removing a divergent factor by 
the normal ordering of the operator, the resulting term is proportional 
to $g^{MN}$ and we can identify it as coming from
the dilaton part of $V(z)$. (The contraction of $\partial X^i$ or 
$\bar{\partial} X^j$ with $e^{ikX}$ using (\ref{firsttwo})
generates other terms. However they 
vanish because of the 
physical state condition $h_{MN}g^{NN'} k_{N'}=0$.)

Now let us turn to the graviton part of $V(z)$. We are
interested in the case 
when the graviton is polarized in the noncommutative directions.
As we did in the previous subsection, let us start with a warm-up
exercise by paying attention to 
the $e^{ipX}$ part of the open string vertex operators. 
Since the first two terms in (\ref{prop}) contribute
only to the dilaton part of $V(z)$ and the third term
is of the order $O(\alpha')$,  
we can compute the contraction of $V(z)$ with $e^{ipX}$
using only the fourth term in the propagator (\ref{prop}),
\eqn{reducedpropagator}
{ \langle X^i(z) X^j(w) \rangle_{the~fourth} =
  -{1 \over 2\pi } \theta^{ij} \log\left( {z-\bar{w} \over \bar{z} - w}
\right). }
The result is
\ber \label{gravitachyon}
&&(z-\bar{z})^2
\langle \partial X^i \bar{\partial} X^j e^{ikX(z)}
\prod_{a=1}^n e^{ip_aX(t_a)} \rangle \nonumber\\
&=&(z-\bar{z})^2   \exp\left[-{i\over 2} 
\sum_{a<b} p_{a}\theta p_{b}\epsilon(t_a-t_b)
+ \sum_a i  k \theta p_a \tau(t_a,z)\right]
 \delta(k+p_1+\cdots+p_n)\times \nonumber \\
&&~~~~~~~~\times\sum_{a,b=1}^n {1\over 2\pi i}\left( 
{(\theta k)^i \over z-\bar{z}} + \sum_{a=1}^n
{(\theta p_a)^i \over z-t_a}\right)
 {1\over 2\pi i}\left( 
{(\theta k)^j \over z-\bar{z}} - \sum_{b=1}^n
 {(\theta p_b)^j \over \bar{z}-t_b}\right)
\nonumber\\
&=&  \exp \left[-{i\over 2} 
\sum_{a<b} p_{a}\theta p_{b}\epsilon(t_a-t_b)
+ \sum_a i  k \theta p_a \tau(t_a,z)\right]
 \delta(k+p_1+\cdots+p_n)
\times
\nonumber \\
&& ~~~~~~~~~~~~~\times {1\over (2\pi i)^2}
\sum_{a,b=1}^n 
   e^{-2\pi i \tau(t_a, z)+2\pi i \tau(t_b, z)}
(\theta p_{a})^i  
 (\theta p_{b})^j.
\eer
To go from the second to the third line, we used
the momentum conservation $k = - (p_1 + \cdots + p_n)$ and
the identities,
\eqn{identity}
{ {1 \over z-\bar{z}} - {1 \over z-t}
 = {1\over z-\bar{z}} e^{-2\pi i \tau(z,t)}~,~~~
{1 \over z-\bar{z}} + {1 \over \bar{z}-t}
 = {1\over z-\bar{z}} e^{+2\pi i \tau(z,t)}.}
The inverse Fourier transform of this then gives 
\ber \label{gravitachyoncoord}
&&  (z-\bar{z})^2 \langle \partial X^i \bar{\partial} X^j e^{ikX(z)}
\prod_{a=1}^n
f_a(X(t_a))  \rangle \nonumber\\
&=&{1 \over (2\pi)^2}  \int dx~ *\left[
~e^{ikx} 
\sum_{a, b=1}^n e^{-2\pi i\tau(t_a,z)+2\pi i \tau(t_b, z)} 
\theta^{im}\theta^{jn}
  {\partial^2 \over \partial x_a^{m} \partial x_b^{n}}
\prod_{c=1}^n f_c(x_c) \right],
\eer            
where we set
\eqn{whatxa}
{x_a^i = x^i + l^i \tau(t_a,z),~~~~a=1,\ldots,n,}
after taking the derivatives on the right-hand
side of (\ref{gravitachyoncoord}). 
We see that the one-point function of the graviton
vertex operator is given by acting the 
second order differential operator,
\eqn{gravitydifferential}
{\sum_{a, b} e^{-2\pi i\tau(t_a,z)+2\pi i \tau(t_b, z)} 
\theta^{im}\theta^{jn} 
{\partial^2 \over \partial x_a^{m} \partial x_b^{m}},}
on the corresponding formula (\ref{coord}) for
the one-point function of the close string tachyon.

We are now ready to discuss the amplitudes involving 
the gauge fields $A_{i}, A_\mu$
and the scalar fields $\Phi^\alpha$. As one
can expect, we need to replace the partial
derivatives in (\ref{gravitydifferential}) with the
covariant derivatives as in 
\ber 
&& F_{ij} =
\partial_i A_j - \partial_j A_i + i (A_i*A_j - A_j*A_i), \nonumber\\ 
&&D_i \Phi^\alpha = \partial_i \Phi^\alpha + i (A_i* \Phi^\alpha
- \Phi^\alpha * A_i).
\eer
The nonlinear terms in the covariant derivatives
arise from integration by parts in
the integrals over the locations $t_a$ of the open
string vertex operators, taking into account the fact 
that we are using the point splitting regularization
on the worldsheet. 
As in the case of the closed string tachyon state discussed
in the previous subsection, it is straightforward to
turn this into the string theory $S$-matrix computation
and to read off the coupling of the graviton to the
gauge fields and the scalar fields on the branes. 
We find that the bulk graviton $h_{ij}$ polarized in the
noncommutative directions  couples to the gauge theory
operator $T^{ij}$
given by 
\ber \label{emforkneq}
&&T^{ij} = 
{\theta^{ii'}\theta^{jj'} +\theta^{ji'}\theta^{ij'}\over
(2\pi)^2}
 \int dx* \left[e^{ikx} 
\exp \left( i\int_0^1 d\tau ~ l^i
A_i(x+l\tau)  + y_\alpha \Phi^\alpha(x+l \tau) \right)\times \right.
\nonumber \\
&&~~~~~~\times \left\{
 i\int_0^1 d\tau e^{-2\pi i \tau}\left( l^{m} F_{i'm}(x+l\tau) 
 + y_\alpha D_{i'} \Phi^\alpha(x+l\tau) \right)\times \right. \nonumber\\
&&~~~~~~~~~~~\times
 i\int_0^1 d\tau' e^{+2\pi i \tau'}\left( l^{n} F_{j'n}(x+l\tau') 
 + y_\beta D_{j'} \Phi^\beta(x+l\tau') \right)  ~+\nonumber\\
&&~~~~~~~~~~~~~~~~~~~\left.\left. + ~
 i\int_0^1 d\tau\Big(  l^{m} D_{i'} F_{j'm}(x+l\tau) + y_\alpha D_{i'}
               D_{j'} \Phi^\alpha(x+l\tau) \Big)\right\}\right].
\eer
A detailed derivation of this formula is presented in Appendix A.

There are several interesting aspects of this operator $T^{ij}$.
It has been expected in the literature that the operators
in the noncommutative gauge theory dual to particle states 
in the bulk take the form,
\eqn{conjecture}
{ \widehat{{\cal O}}(k) = \int dx  * 
\left[e^{ikx}~ {\rm STr}\prod_i \int_0^1 d\tau~ B_i(x+l\tau) 
\exp\left( i\int_0^1 d\tau ~l^i A_i(x+l\tau) \right)
\right],}
when the corresponding operators in the theory in commutative
space are given by 
\eqn{commutative}
{ {\cal O}(k) = \int dx~ e^{ikx}~ {\rm STr} \prod_i B_i(x).}
(Here we have written the symmetrized trace ${\rm STr}$ 
over the gauge group index explicitly.)  
The expression (\ref{emforkneq}) we found for
$T^{ij}$ does not follow this prescription. 
One of the motivation for the conjecture (\ref{conjecture})
was that it is the simplest solution to the requirement
that $\widehat{{\cal O}}(k)$ reduces to ${\cal O}(k)$
in the commutative limit, 
\eqn{commlimit}
{\widehat{{\cal O}}(k) \rightarrow {\cal O}(k),~~~
{\rm as}~~\theta^{ij} \rightarrow 0.}
The operator $T^{ij}$ 
we have derived {\it does not
satisfy this requirement}. Instead it vanishes
in this limit because of the factor 
$(\theta^{ii'}\theta^{jj'}+\theta^{ij'}\theta^{ji'})$ 
in front of everything.  

This result may seem paradoxical since 
the worldsheet propagator (\ref{prop})
we have used to derive (\ref{emforkneq})
reduces to the one for the commutative
case if we substitute $\theta^{ij} =0$
and $g^{ij} = G^{ij}$. To clarify the situation,
it is useful to point out that the amplitude (\ref{gravitachyoncoord})
and the corresponding formula for the gauge
fields computed in Appendix A do not contain
any factor of $\alpha'$. These amplitudes
in the noncommutative case are non-zero and finite 
in the zero slope limit. It is because they
are computed using the propagator 
(\ref{reducedpropagator}), which does not contain
$\alpha'$.  This is to be contrasted
to the commutative case when 
the standard expression
for the energy-momentum tensor 
\eqn{commuem}
{T_{commutative}^{ij} \sim F^i_{~k} F^{kj}
+ F^i_{~\mu} F^{\mu j} 
  - {1\over 4} G^{ij} F^2 + \cdots,}
is obtained from $O(\alpha'^3)$ terms 
in the disk amplitude of one graviton
and several gauge fields. The amplitude that leads
to (\ref{commuem}) 
involves at least 3 propagators on the worldsheet, 
each of which costs one power of $\alpha'$. 
The fact that $T^{ij}$ in the noncommutative 
case vanishes in the limit
$\theta^{ij} \rightarrow 0$ is consistent
with the fact that there is no $O(1)$ terms
in the commutative case. This resolves the
apparent paradox.

Another interesting aspect of the energy-momentum
tensor $T^{MN}$ is the way it is conserved,
\eqn{conserved}
{k_M T^{MN} = 0.}
Let us first examine the conservation of $T^{Mj}$,
which involves $T^{ij}$, $T^{\mu j}$ and $T^{\alpha j}$. 
One can show that the leading 
terms in the mixed components $T^{\mu j}$
and $T^{\alpha j}$ are of the order $O(\alpha')$. 
Since $k_i, k_\alpha \sim O(1)$ and
$k_\alpha \sim O(\alpha'^{-1})$, 
the conservation law at the zero slope limit is 
\eqn{limitconserved}
{k_i T^{ij} + k_\alpha T^{\alpha j} 
 = k_i T^{ij} + {y_\alpha\over 2\pi \alpha'} T^{\alpha j} = 0.}
The expression for $T^{\alpha j}$ is 
\ber \label{talphaj}
 &&T^{\alpha j}
 = \alpha' \theta^{jm}
 \int dx~* \left[e^{ikx}~ 
\exp \left( i\int_0^1 d\tau ~ l^i
A_i(x+l\tau) + y_\alpha \Phi^\alpha(x+l \tau) \right)\times \right.
\nonumber \\
&&\times \left\{
 \int_0^1 d\tau e^{-2\pi i \tau}  \Phi^\alpha(x+l\tau) 
 \int_0^1 d\tau' e^{+2\pi i \tau'}
i\left( l^n F_{mn}(x+l\tau') 
 + y_\beta D_{m} \Phi^\beta(x+l\tau') \right) \right. \nonumber\\
&&\left.\left.-
 \int_0^1 d\tau e^{+2\pi i \tau} \Phi^\alpha(x+l\tau) 
 \int_0^1 d\tau' e^{-2\pi i \tau'}
i\left( l^n F_{mn}(x+l\tau') 
 + y_\beta D_{m} \Phi^\beta(x+l\tau') \right)\right\}\right].\nonumber \\
\eer
Curiously, we do not have to use the equation
of motion of the gauge fields and the scalar fields
in order to prove the conservation law (\ref{limitconserved}).  
It just follows from the kinematical identities such as
\ber \label{kinematics}
&& k_i \theta^{ii'} l^{m} F_{i'm}  =  
 l^i l^j F_{ij} = 0,\nonumber \\
&& k_i \theta^{ii'} \int_0^1 d\tau  \partial_{i'} f(x+l\tau)
 =\int_0^1 d\tau \partial_\tau f(x+l\tau)
= f(x+l) - f(x).
\eer
The first line in (\ref{kinematics}) vanished because of the 
antisymmetry of $F_{ij}$. 
The right-hand side of the second line vanishes 
when it is inserted in $\int dx * [e^{ikx} \cdots ]$,
taking into account the path-ordering along the Wilson line. 
A detailed proof of the conservation law (\ref{limitconserved})
is given in Appendix B. To describe salient features
of the proof, let us look at $e^{-2\pi i \tau}
\theta^{ii'}(l^m F_{i' m} + y_\alpha D_{i'} \Phi^\alpha)$
in the second line of (\ref{emforkneq}). If
we multiply $k_i$ to it, the first term 
vanishes because of the first identity in (\ref{kinematics}).
As for the second term, 
we can use the second identity in (\ref{kinematics})
to perform the integration by parts. Because of the
factor $e^{-2\pi i\tau}$,
we obtain an additional term of the form
$2\pi i \Phi^\alpha$ from the integration by parts.
It is canceled by $y_\alpha T^{\alpha j}$
because of the second line in (\ref{talphaj})
multiplied by $y_\alpha$. 

Similarly one can prove the conservation of $T^{M \beta}$,
\eqn{anotherconserve}
{k_i T^{i\beta} + k_\alpha T^{\alpha\beta} 
= k_i T^{i\alpha} + {y_\alpha \over 2\pi \alpha'}
T^{\alpha\beta} = 0,}
where $T^{\alpha\beta}$ is
\ber \label{talphabeta}
 &&T^{\alpha \beta}
 = -(2\pi \alpha')^2 
 \int dx~* \left[e^{ikx}
\exp \left( i\int_0^1 d\tau~ l^i
A_i(x+l\tau) + y_\alpha \Phi^\alpha(x+l \tau) \right)\times \right.
\nonumber \\
&&~~~\times \left. \left\{
 \int_0^1 d\tau e^{-2\pi i \tau}  \Phi^\alpha(x+l\tau) 
 \int_0^1 d\tau' e^{+2\pi i \tau'} \Phi^\beta(x+l\tau')
+ (\alpha \leftrightarrow \beta) \right\} \right].
\eer

Let us discuss implications of the fact that the
energy-momentum tensor is kinematically conserved.
It is well-known that, given a symmetry of a Lagrangian,  
the  construction of the corresponding N\"other current
is not unique since one can add a term of the
form $\partial_j J^{ij}$ where $J^{ij}$ is
anti-symmetric in $i$ and $j$. Such a term 
is conserved without using any equation of motion.
This ambiguity has a physical origin; 
when the current is coupled to a gauge field $a_i$, the term
$\partial_j J^{ij}$ represents the coupling
to the field strength,
\eqn{improvedcoupling}
{  (\partial_i a_j - \partial_j a_i) J^{ij}.}
One can say that $\partial_j J^{ij}$ is conserved
without using the equation of motion since the term
(\ref{improvedcoupling}) does not modify the equation
of motion when the gauge field is flat. 
A similar term can arise in the energy-momentum
tensor when the matter couples directly to the 
curvature of the metric. Such a coupling does not modify
the equation of motion in flat
space but changes the form of the energy-momentum 
tensor. An example of this is found in the Coulomb-Gas
formalism of two-dimensional conformal field theory
where one modifies the energy-momentum
tensor of the massless free scalar field $\varphi$ as
\ber \label{coulombgas}
 T_{zz} &=& {1\over 2} \partial \varphi^2  
~~~\rightarrow~~~ {1\over 2} \partial \varphi^2 + \kappa
 \partial^2 \varphi, \nonumber\\
T_{z\bar{z}} &=& 0 ~~~~~~~~ \rightarrow ~~~~
 - \kappa \partial \bar{\partial} \varphi,
\eer
with some constant $\kappa$. These additional terms have
physical significance. For example, the conformal
anomaly of the theory depends on $\kappa$.  

The fact that the energy-momentum tensor $(T^{ij},
T^{\alpha j} )$ presented in (\ref{emforkneq})
and (\ref{talphaj}) is conserved without using
the equation of motion suggests that there is
a coupling of the bulk metric to the noncommutative
gauge theory which vanishes in the limit of 
flat space. It is curious that such a coupling
appears in the leading order in the $\alpha'$ expansion.
It would be interesting to compute amplitudes
involving more gravitons and to identify the 
precise form of the coupling. 

\subsection{Graviton in Superstring} 

To compute the disk amplitude of one graviton
and many gauge fields in 
superstring theory, it is convenient to
use the operator in the $(-1,-1)$-picture for the graviton,
\eqn{supergraviton}
{ V^{(-1,-1)}(z) = {1\over 2} \delta(\gamma) \delta(\bar{\gamma})
h_{MN}(k) \psi^M(z) \bar{\psi}^N(\bar{z}) e^{ikX(z)},}
and the operator in the $0$-picture for
open string,
\ber \label{superopen}
U^{(0)}(t) & = &u_i(p)\left( {dX^i \over dt} -2i
 p \cdot \Psi ~\Psi^i\right)e^{ipX}
    + u_\mu(p) \left( {dX^\mu \over dt} -2i
              p \cdot \Psi~ \Psi^\mu\right)e^{ipX} +
  \nonumber \\
&& ~~~~~~~~~~~~~~~~ + 
     \phi^\alpha (p) g_{\alpha\beta}
        \left( i \partial_\perp X^\beta
       - 2i p\cdot \Psi ~\Psi^\beta\right)e^{ipX},
\eer
where $\gamma$ and $\bar{\gamma}$ are bosonic
ghosts, $\psi$ and $\bar{\psi}$ are worldsheet fermions
in the left and the right movers, and $p \cdot
\Psi = p_i \Psi^i + p_\mu \Psi^\mu$. In the noncommutative
directions, the fermion propagators are given by
\ber \label{fermionprop}
\langle \psi^i(z) \psi^j(w) \rangle
&=& {\alpha' \over z-w}g^{ij} , \nonumber \\
\langle \psi^i(z) \bar{\psi}^j(\bar{w})
\rangle &=& {\alpha' \over z-\bar{w}}\left(
-g^{ij} + 2G^{ij} + 2 {\theta^{ij}\over 2\pi \alpha'}\right),
\nonumber \\
\langle \bar{\psi}^i(\bar{z})
\psi^j(w) \rangle &=& {\alpha' \over \bar{z} - w}
\left(-g^{ij} + 2G^{ij} - 2 {\theta^{ij} \over 2\pi \alpha'}
\right), \nonumber \\
\langle\bar{\psi}^i(\bar{z})\bar{\psi}^j(\bar{w})
\rangle &=& {\alpha' \over \bar{z} - \bar{w}}g^{ij},
\eer
and the fermions on the boundary are defined by
\ber \label{boundaryfermion}
\Psi^i &=& {1\over 2} (\psi^i + \bar{\psi}^i)~,~~~~
\Psi^\mu = {1 \over 2} (\psi^\mu + \bar{\psi}^\mu), \nonumber\\
\Psi^\alpha &=& {1 \over 2}
(\psi^\alpha - \bar{\psi}^\alpha).
\eer 
The bulk-boundary propagators are 
\ber \label{fermibulkboundary}
\langle \psi^i(z) \Psi^j(t) \rangle &=&
{\alpha'\over z-t}\left(G^{ij} + {\theta^{ij}
\over 2\pi \alpha'}\right) , \nonumber \\
\langle \bar{\psi}^i(\bar{z}) \Psi^j(t) \rangle
&=& {\alpha' \over \bar{z}-t}
\left(G^{ij} - {\theta^{ij}\over 2\pi \alpha'}\right).
\eer
Using these, we can show in the leading order in
the $\alpha'$ expansion that 
\ber \label{fermoingeneratingfun}
&&  (z-\bar{z}) \langle
 {1\over 2}(\psi^{i} \bar{\psi}^{j}
+ \psi^j \bar{\psi}^i) e^{ikX(z)}
 \exp \left( i\int_{-\infty}^\infty
dt ~ A_i{dX^i \over dt}
  - F_{ij} \Psi^i \Psi^j \right)\rangle \nonumber \\
&=&2 \theta^{ii'}\theta^{jj'}
\int dx * \left[ e^{ikx} 
\exp \left( i\int_0^1 d\tau~ l^i
A_i(x+l\tau)\right) \times \right. \nonumber \\
&&~~\times \left. \alpha' G^{mn}
\int_0^1 d\tau (F_{i'm}(x+l\tau)-\theta^{-1}_{i'm})\int_0^1 d\tau'
(F_{j'n}(x+l\tau')-\theta^{-1}_{j'n}) \right] .
\eer
As usual we have removed contributions from the
dilaton part of the closed string vertex operator.
The nonlinear terms in the field strength $F_{ij}$ on
the right-hand side are generated as a consequence
of the point-splitting regularization on the left-hand 
side [10]. 
Similar expressions can be derived for amplitudes
involving $A_\mu$ and $\Phi^\alpha$.

As we did in the previous two subsections, we can
convert (\ref{fermoingeneratingfun}) into the superstring
$S$-matrix computation using,
\eqn{ghostcontraction}
{\langle \delta(\gamma(z)) \delta(\bar{\gamma}(\bar{z}))\rangle
= {1\over z-\bar{z}},}
and 
\eqn{fermionoicghostagain}{
\langle c(z) \bar{c}(\bar{z}) c(t_1) \rangle
 = (z-\bar{z})(z-t_1)(\bar{z}-t_1),}
and extract the gauge theory
operator which couples  to the bulk graviton. The
result is 
\ber \label{susyem}
 &&T^{ij}  = 2 \alpha' \theta^{ii'}\theta^{jj'}
\int dx * \left[e^{ikx}
\exp \left( i\int_0^1 d\tau~ l^i
A_i(x+l\tau) + y_\alpha \Phi^\alpha(x+l \tau) \right)\times \right.
\nonumber \\
&& ~\times \int_0^1d\tau \int_0^1d\tau' \left( G^{mn}
(F_{i'm}(x+l\tau)-\theta^{-1}_{i'm})(F_{j'n}(x+l\tau')
-\theta^{-1}_{j'n}) +\right. \nonumber\\
&&~~\left.\left.
+G^{\mu\nu} F_{i'\mu}(x+l\tau) F_{j'\nu}(x+l\tau') +
    g_{\alpha\beta} D_{i'} \Phi^\alpha(x+l\tau) D_{j'}
\Phi^\beta(x+l\tau') \right)\right].
\eer

The way the energy-momentum tensor is conserved is again
very interesting. Consider $k_iT^{ij}$ and pay 
attention to the second line of (\ref{susyem}).
The multiplication of $k_i$ generates
a term of the form $\int_0^1 d\tau~ l^i F_{im}(z+l\tau)$,
and this can be written as $D_m$ acting on the $A_i$ 
part of the Wilson
line, $\exp(i\int_0^1 d\tau ~l^i A_i(x+l\tau))$. 
We can then perform integration by parts in $D_m$.
This gives rise to two terms. One is 
of the form $k_m G^{mn} F_{j'n}$, which is generated
by acting $D_m$ on $e^{ikx}$. It is cancelled by the terms
in the second line of (\ref{susyem}) containing
$\theta^{-1}$. The other is
of the form $D_m G^{mn} F_{j'n}$ and makes 
a part of the equation of motion for the gauge
fields. A more careful analysis shows that 
$k_M T^{Mj}$ vanishes because of the equation 
of motion. 

The fact that the conservation of the energy-momentum
requires the equation of motion for the gauge fields
means that, unlike the case of bosonic string, 
the coupling of the bulk metric to the
gauge fields does not disappear in the limit
of flat space. 
In fact the expression for the energy-momentum tensor
(\ref{susyem}) is consistent with the way the bulk
metric appears in action for the noncommutative
gauge theory derived in \cite{Seiberg:1999vs},
\ber \label{action}
S & = & {-1 \over g_{YM}^2}\int dx
\sqrt{\det G}* \left[
  ~{1\over 4} G^{MN} G^{PQ} (F_{MP}-\theta^{-1}_{MP})
(F_{NQ}-\theta^{-1}_{NQ})  +\right.\nonumber \\
&&~~~~~~~~~~~~~\left.+
{1 \over 2} G^{MN} g_{\alpha\beta}
D_M \Phi^\alpha  D_N \Phi^\beta
- {1 \over 4} g_{\alpha\beta}g_{\gamma\delta}
 [\Phi^{\alpha}, \Phi^{\gamma}]
 [\Phi^{\beta}, \Phi^\delta] \right] .
\eer 
Here $\theta^{-1}_{MN}=0$ unless $(M,N)=(i,j)$. 
The action
depends on the closed string metric $g_{ij}$
through the open string metric 
\eqn{closedmetric}{G^{ij} ={1 \over (2\pi \alpha')^2} 
 \theta^{im} \theta^{jm} g_{nm},}
and the gauge coupling constant 
\eqn{gaugecoupling}{
g_{YM}^2 = {2\pi g_{{\rm string}}\over
   \sqrt{\det \left({
\theta^{im}g_{mj}\over 2\pi \alpha'}\right)}},}
where $g_{\rm string}$ is the string coupling constant.
A variation of the action with respect to $g_{ij}$
gives
\ber \label{metricvariation}
 {\partial S \over \partial g_{ij}}
 &=& {-\theta^{ii'}\theta^{jj'}\over 2(2\pi\alpha')^2 g_{YM}^2} 
\int dx \sqrt{\det G} 
* \left[ G^{mn} (F_{i'm}-\theta^{-1}_{i'm})(
F_{j'n}-\theta^{-1}_{j'n}) +\right. \nonumber\\
& & ~~~~~~~~~~~~~~~~~~~~~~~~~~\left. + G^{\mu\nu} F_{i'\mu}
         F_{j'\nu}+ 
g_{\alpha\beta}D_{i'} \Phi^\alpha D_{j'}
\Phi^\beta \right] .
\eer
To derive this, we used the fact that the combination
$\sqrt{\det G}/g_{YM}^2$ is independent of the closed
string metric $g_{ij}$. Since the action (\ref{action})
was derived assuming the bulk metric is flat,
we can only make the variation with respect to
constant $g_{ij}$.
It is satisfactory to see that 
our result (\ref{susyem}) agrees with
(\ref{metricvariation}) in the limit 
$k \rightarrow 0$. 

It is important to point out that the expression
for the energy-momentum tensor (\ref{susyem})
still does not reduce to the one
in the commutative case (\ref{commuem}) 
in the limit $\theta^{ij}\rightarrow 0$. 
In the string theory computation, 
(\ref{susyem}) comes from the order $\alpha'$
terms of the disk amplitude whereas
the commutative result is from the order $\alpha'^3$. 
The difference in the powers of $\alpha'$ in these 
computations is manifest in the factor $\alpha'^{-2}$
on the right-hand side of (\ref{metricvariation}). 

\subsection{Massive String Modes}

Disk amplitudes for massive modes of close string 
can be evaluated in the same way as we 
did for the tachyon and the massless modes.
The basic ingredient is the open Wilson line
connecting $x$ and $x+l$ with the condition 
\eqn{onshellagain}{
  G_{ij} l^i l^j  + g^{\alpha \beta} y_\alpha
y_\beta = 0.}
As we discussed earlier, this condition
is independent of 
the amount ${\bf N}$ of closed string excitations. 
Each $\partial X^i$ in the closed string 
vertex operator $V(z)$ is involved in the 
self-contraction within $V(z)$ 
(as in the case of the dilaton) or introduces
the gauge theory operator 
$e^{-2\pi i \tau}\theta^{ij} ( l^m F_{jm}(x+l\tau)  
+ \cdots)$ integrated along the Wilson line
(as in the case of the graviton). More
derivatives on $\partial X^i$ induce 
more derivatives on $F_{jm}$, $etc$.  

\section{Comments on Gravity Dual} 

In this paper, we studied coupling of
noncommutative gauge theories on branes to close string
states in the bulk. We showed that a basic ingredient
is the straight open Wilson line. It may seem paradoxical 
that a {\it closed} string can couple to something 
that is {\it open}. This is possible since 
the distance between the two end-point of the
Wilson line measured by the closed string metric,
$\sqrt{g_{ij}l^i l^j}$, is of the order $\alpha'$,
which is much smaller than the string length $\sqrt{\alpha'}$. 
Therefore, closed string states of the size of 
$\sqrt{\alpha'}$ in the bulk can easily couple
to the open Wilson line of the length $\sim \alpha'$.  

There are several large $N$ noncommutative gauge 
theories that are expected to have dual gravitational
descriptions, generalizing the $AdS$/CFT correspondence 
\cite{Maldacena:1998re}
to the case when the NS-NS 
two-form field is non-trivial \cite{Hashimoto:1999ut,
Maldacena:1999mh,oz}. For the four-dimensional 
${\cal N}$=4 gauge theory with two noncommutative directions, the
metric of the dual gravitational theory is \cite{Hashimoto:1999ut,
Maldacena:1999mh}
\ber \label{adscft}
{ds^2 \over \sqrt{\lambda}\alpha'}
& =&{du^2 \over u^2} + u^2 G_{\mu\nu} dx^\mu dx^\nu
  + {u^2 \over 1 + \lambda \theta^2 u^4}G_{ij} dx^i dx^j + d\Omega_5^2,
\nonumber\\
&&~~~~~~~~~~\lambda = g_{YM}^2 N,
\eer
where the metric is given in the string frame,
$\mu,\nu=0,1$ and $i,j=2,3$ with $\theta=\theta^{23}$, and $d\Omega_5^2$
is the metric on the unit 5-sphere. Near the boundary
$u \rightarrow \infty$, the metric in the commutative directions 
$(x_0, x_1)$ behaves as in $AdS$. Relative to this, 
the metric in the noncommutative directions 
$(x_2,x_3)$ scales as $u^{-4}$. This scaling is
equivalent to $g_{ij} \sim \alpha'^2$ used throughout this paper if we
identify $u$ as the UV cutoff scale of the gauge theory. 
Therefore, for the same reason as in the previous paragraph,
 particles states in the bulk can couple
to the open Wilson line located at the boundary of
the geometry (\ref{adscft}). It is not necessary 
to look for a classical string worldsheet 
ending on the Wilson line at the boundary; such a description 
breaks down for large $u$ when the size of the Wilson 
line measured by the bulk metric becomes smaller than the string 
length. 

There has been evidence that correlation 
functions of supergravity fields computed in 
the geometry (\ref{adscft}) agree with gauge 
theory expectations \cite{Gross:2000ba,Rozali:2000np}. 
To make a quantitative comparison between 
the gauge theory and its gravity dual, we need 
to know a precise map between gauge theory operators 
and supergravity fields. We believe the present 
paper provides a framework to answer this question. 

The gauge theory operators we constructed are
in fixed momentum states in the noncommutative
directions but can be localized in the commutative
directions along the brane. This is because the
momentum $k_\mu$ in the commutative directions
is not constrained by the on-shell condition 
(\ref{reducedonshell}). Moreover, the operators are parametrized by
$y_\alpha$, which is a vector in the transverse
directions. In the case of the four-dimensional 
${\cal N}$=4 theory, $y_\alpha$ is a 6-dimensional
vector obeying the on-shell condition\footnote{Here
we made the replacement $y_\alpha \rightarrow i y_\alpha$
as we discussed in the footnote 2 in Section 2.}
\eqn{newconst}
{ g^{\alpha\beta} y_\alpha y_\beta = 
G_{ij} \theta^{im}\theta^{jn}  k_m k_n.}
For a given $k_i$, we may regard $y_\alpha$ as 
coordinates on $S^5$. By expanding the Wilson line
in powers of $y_\alpha$, we generate infinitely
many representations of the $SO(6)$ $R$ symmetry. 
This is to be contrasted to the case of the
theory on commutative space. There operators
are classified into irreducible representations
of the $R$ symmetry since operators with different
$R$ charges are renormalized differently 
and have different conformal weights. 

We would like to point out that the construction
of Wilson lines in the noncommutative theory
fits well with its dual gravitational description. 
To see how it works, it would be useful to remind ourselves
of the situation of the commutative case. There
the gravity dual is the string theory on $AdS_5 \times S^5$
with the metric, 
\eqn{ads5s5}
{ {ds^2 \over \sqrt{g_{YM}^2 N} \alpha'}
= {du^2 \over u^2} + u^2 (-dx_0^2 + dx_1^2 +dx_2^2+dx_3^2)
+ d\Omega_5^2.}
Consider the Klein-Gordon equation in this geometry
for a scalar field with mass $m$. 
For an eigenstate of the momentum in the $x$ direction 
and the angular momentum
in the $S^5$ direction, the equation becomes
\eqn{kgequation}
{ \left( -u^2{d^2 \over du^2}  + \tilde{m}^2 
    + j(j+4) + {15\over 4}
+ {k^2 \over u^2}   
\right) u^{5/2} \Psi (u) = 0,}
where $j$ is the orbital angular momentum of $SO(6)$,
which we regard as the $R$ charge, 
and $\tilde{m}^2 = m^2 \sqrt{g_{YM}^2 N}\alpha'$. 
For large $u$, a solution to this behaves as
\eqn{largeuads}
{ \Psi(u) \sim u^{\Delta_\pm - 4}
                \left( 1 + O\left({1\over u}\right)\right),
~~~~~u\rightarrow \infty,}
where
\eqn{whatdelta}
{\Delta_\pm = 2 \pm \sqrt{\tilde{m}^2 + (j+2)^2}. }
It is important to note that the exponent of $u$ depends on 
the $R$ charge $j$ as well as the mass $\tilde{m}$. 
Thus operators with different $j$
scale differently as $u \rightarrow \infty$.
On the other hand, since the momentum $k$ does
not appear in the scaling behavior in (\ref{largeuads}), 
we can take the inverse
Fourier transformation with respect to $k$ and
consider a source localized at a point $x_0$
on the boundary. Namely, we can impose 
a boundary condition on the scalar field $\Psi$ so that 
it behaves near the boundary as
\eqn{bccommutative}
{\Psi \sim u^{\Delta_+ - 4} 
\delta^{(4)}(x-x_0),~~~~u\rightarrow \infty.}
The exponent of $u$ is then related to
the conformal weight of the corresponding operator
in the gauge theory inserted at $x_0$.  This 
gives the correspondence between operators in
the gauge theory and boundary conditions on
the bulk fields \cite{Maldacena:1998re,
Gubser:1998bc,Witten:1998qj}.  

Now let us turn to the geometry (\ref{adscft})
for the noncommutative gauge theory. In this background,
the dilaton $\phi$ also depends on $u$,
\eqn{whatdilaton}
{e^{2\phi} = {\lambda \over N} 
{1 \over 1 + \lambda\theta^2 u^4}.}
For definiteness, we consider a massive scalar field 
minimally coupled
to the metric in the Einstein frame, $g_{MN}^E =
e^{-\phi/2} g_{MN}$. The equation corresponding to 
(\ref{kgequation}) is then
\ber \label{kgnoncommutative}
&&\left( - u^2 {d^2 \over du^2}
+u^2 \lambda G_{ij} l^i l^j + \tilde{m}^2 
(1 + \lambda \theta^2 u^4)^{1\over 4} 
 +\right. \nonumber \\
&&~~~~~~~~ \left. + j(j+4) + {15 \over 4} + 
{G^{\mu\nu} k_\mu k_\nu + G^{ij} k_i k_j \over u^2}
  \right) u^{5/2} \Psi(u) = 0,
\eer
where $l^i = k_j \theta^{ji}$ as usual and 
$\tilde{m}$ is the mass with some rescaling. 
For large $u$, a solution to this equation is
\eqn{largeunoncomm}
{ \Psi(u) = u^{\lambda_\pm} e^{
 \pm u \sqrt{\lambda G_{ij} l^i l^j}}\left(
1 + O\left( {1 \over u} \right) \right),}
where
\eqn{whatlambdapm}{
\lambda_\pm = 
 -{5 \over 2} \pm {\tilde{m} \theta^{1/2}
\over 2\sqrt{\sqrt{\lambda}G_{ij} l^i l^j}}.}
The $u$-dependent factor on the right-hand side
are to be removed by the renormalization of the 
corresponding operator.\footnote{The linear term in the exponent
$\pm \sqrt{\lambda G_{ij}l^i l^j} u$ means that $\Psi(u)$
has an imaginary momentum in the $u$ direction.
This fits well with our discussion at the beginning
of Section 2 and in the footnote 2.}
We note that these terms are independent of
the $R$ charge $j$ as well as the momentum $k_\mu$
in the commutative directions along the brane.
Therefore, we can perform the inverse Fourier transformation
on them and construct
a solution which grows exponentially toward a
point $\Omega_0$ on $S^5$ and $x_0$ in the commutative 
directions on the boundary.
We have found that the wave equation in the geometry
(\ref{adscft}) allows us to impose the boundary condition
\eqn{bulkboundarycondition}
{ \Psi \sim u^{\lambda_+} e^{
 u \sqrt{\lambda G_{ij} l^i l^j}}
\delta^{(2)}(x - x_0) \delta^{(5)}(\Omega-\Omega_0),
~~~~u\rightarrow \infty.}
It is natural to interpret it as a boundary condition
corresponding to an insertion of a gauge theory
operator at the point $x_0$ in the commutative direction.
The point $\Omega_0$ on $S^5$ specifies the parameter
$y_\alpha$ in the Wilson line. Given the boundary
condition (\ref{bulkboundarycondition}), one can 
solve the wave equation in the bulk, compute 
a coefficient in front of the exponentially decaying part 
$\sim e^{-u\sqrt{\lambda G_{ij}l^il^j}}$
of the solution and extract information
on the strongly coupled dynamics of the
noncommutative gauge theory as discussed
in \cite{Hashimoto:1999ut,Maldacena:1999mh,Gross:2000ba}.

We have found that, in the noncommutative case, 
it is possible to impose a boundary condition 
on a particle state in the bulk that it ends on 
a point on $S^5$ as well as on a point in
the commutative directions. 
This is to be contrasted to the commutative
case when point particle states
in $AdS_5 \times S^5$ are classified into irreducible 
representations of $SO(6)$.
The boundary condition in the noncommutative
case reminds us of the one for Wilson loop in
the gravitational dual of the ${\cal N}$=4 theory
in commutative space \cite{Rey:1998ik, Maldacena:1998im}.
There a loop is parametrized
by $x^\mu(t)$ ($\mu = 0, \ldots, 3$) 
and $y_\alpha(t)$ ($\alpha = 1, \ldots, 6$)
obeying the condition \cite{Drukker:1999zq},
\eqn{commloop}
{ G_{\mu\nu} \dot{x}^\mu \dot{x}^\nu 
 + g^{\alpha\beta} \dot{y}_\alpha \dot{y}_\beta
= 0,}
similar to the on-shell condition
(\ref{constraintfirst}) in our case. 
The loop variables $(x(t), y(t))$ 
specify a boundary condition
on a semi-classical string worldsheet on 
$AdS_5 \times S^5$ that it terminates 
at a loop $x(t)$ on the boundary of $AdS_5$ 
and $\dot y(t)$ on $S^5$. In this respect, 
a particle state in the
noncommutative case behaves similarly to
a semi-classical 
worldsheet in the commutative case. 

\bigskip
\centerline{*~~~~~~*~~~~~~*}
\bigskip

In this paper, we developed a method to 
determine gauge theory operators dual to any closed 
string states including massive ones. In particular, 
we found the precise form of the gauge theory operator $T^{ij}$ 
coupled to the bulk graviton polarized in the noncommutative
directions in both bosonic string theory and superstring
theory. 
In either case, the expression is different from
what was expected in the literature.   
In the case of bosonic string, the structure 
of $T^{ij}$ suggests that the leading coupling between
the gauge fields on the brane and the graviton 
in the bulk is through the curvature of the bulk metric. 
In the case of superstring, the expression for $T^{ij}$
is consistent with the way the bulk metric appears in
the action of the noncommutative gauge theory 
\cite{Seiberg:1999vs}.

\bigskip

\section*{Acknowledgments}

We would like to thank Jaume Gomis, Thomas Mehen,
John Schwarz,
Mark Wise and Edward Witten for discussions. 
The research was supported in part by
the DOE grant DE-FG03-92ER40701
and the Caltech Discovery Fund.

\newpage
\appendix
\section{Energy-Momentum Tensor $T^{ij}$}

In this appendix, we give a detailed derivation of
the operator $T^{ij}$ in the case of bosonic string theory.
Let us first consider vertex operators for $A_i$,
\begin{equation}
(z-\bar{z})^2 \langle
\partial X^i \bar{\partial} X^j e^{ikX(z)}
\prod_{a=1}^n u_l (p_a) \frac{dX^l}{dt_a} e^{i p_a X(t_a)}
\rangle.
\end{equation}
There are four types of contractions;\\
1. Both $\partial X^i$ and $\bar{\partial} X^j$ contract
with $e^{ikX(z)}$ or $e^{i p X(t)}$.\\
The computation is essentially the same as (\ref{gravitachyon})
and the result is given by
\begin{eqnarray}
&&  \exp \left[-{i\over 2} 
\sum_{a<b} p_{a}\theta p_{b}\epsilon(t_a-t_b)
+ \sum_a i  k \theta p_a \tau(t_a,z)\right]
 \delta(k+p_1+\cdots+p_n)
\times
\nonumber \\
&& \times
\sum_{a,b=1}^n 
\frac{i (\theta p_{a})^i}{2 \pi}
e^{-2\pi i \tau(t_a, z)}
\frac{i (\theta p_{b})^j}{2 \pi}
e^{2\pi i \tau(t_b, z)}
\prod_{c=1}^n k \theta u_c
\frac{\partial \tau (t_c, z)}{\partial t_c},
\end{eqnarray}
where $u_{c}$ is an abbreviation for $u (p_c)$
and we use the same notation in what follows.\\
2. Both $\partial X^i$ and $\bar{\partial} X^j$ contract
with $dX^{l}/dt$.
\begin{eqnarray}
&&  \exp \left[-{i\over 2} 
\sum_{a<b} p_{a}\theta p_{b}\epsilon(t_a-t_b)
+ \sum_a i  k \theta p_a \tau(t_a,z)\right]
 \delta(k+p_1+\cdots+p_n)
\times
\nonumber \\
&& \times
\sum_{a \ne b}
-i (\theta u_a)^i
e^{-2\pi i \tau(t_a, z)}
\frac{\partial \tau (t_a, z)}{\partial t_a}
i (\theta u_b)^j
e^{2\pi i \tau(t_b, z)}
\frac{\partial \tau (t_b, z)}{\partial t_b}
\times
\nonumber \\
&& \times
\prod_{c \ne a,b} k \theta u_c
\frac{\partial \tau (t_c, z)}{\partial t_c},
\end{eqnarray}
where we used
\begin{equation}
\frac{1}{(z-t)^2}
= \frac{2 \pi i}{z-\bar{z}}
e^{-2\pi i \tau(t, z)}
\frac{\partial \tau (t, z)}{\partial t}, \qquad
\frac{1}{(\bar{z}-t)^2}
= \frac{2 \pi i}{z-\bar{z}}
e^{2\pi i \tau(t, z)}
\frac{\partial \tau (t, z)}{\partial t}.
\end{equation}
3. $\partial X^i$ contracts with $dX^{l}/dt$
and $\bar{\partial} X^j$ contracts
with $e^{ikX(z)}$ or $e^{i p X(t)}$.
\begin{eqnarray}
&&  \exp \left[-{i\over 2} 
\sum_{a<b} p_{a}\theta p_{b}\epsilon(t_a-t_b)
+ \sum_a i  k \theta p_a \tau(t_a,z)\right]
 \delta(k+p_1+\cdots+p_n)
\times
\nonumber \\
&& \times
\sum_{a,b=1}^n 
-i (\theta u_a)^i
e^{-2\pi i \tau(t_a, z)}
\frac{\partial \tau (t_a, z)}{\partial t_a}
\frac{i (\theta p_{b})^j}{2 \pi}
e^{2\pi i \tau(t_b, z)}
\prod_{c \ne a} k \theta u_c
\frac{\partial \tau (t_c, z)}{\partial t_c}.
\end{eqnarray}
4. $\partial X^i$ contracts 
with $e^{ikX(z)}$ or $e^{i p X(t)}$
and $\bar{\partial} X^j$ contracts
with $dX^{l}/dt$.
\begin{eqnarray}
&&  \exp \left[-{i\over 2} 
\sum_{a<b} p_{a}\theta p_{b}\epsilon(t_a-t_b)
+ \sum_a i  k \theta p_a \tau(t_a,z)\right]
 \delta(k+p_1+\cdots+p_n)
\times
\nonumber \\
&& \times
\sum_{a,b=1}^n
\frac{i (\theta p_{a})^i}{2 \pi}
e^{-2\pi i \tau(t_a, z)}
i (\theta u_b)^j
e^{2\pi i \tau(t_b, z)}
\frac{\partial \tau (t_b, z)}{\partial t_b}
\prod_{c \ne b} k \theta u_c
\frac{\partial \tau (t_c, z)}{\partial t_c}.
\end{eqnarray}
These four contractions are combined to give
\begin{eqnarray}
&&  \exp \left[-{i\over 2} 
\sum_{a<b} p_{a}\theta p_{b}\epsilon(t_a-t_b)
\right]
 \delta(k+p_1+\cdots+p_n)
\times
\nonumber \\
&& \times
\Bigg[ \sum_{a \ne b}
\left(
\frac{i (\theta p_{a})^i k \theta u_a}{2 \pi}
-i (\theta u_a)^i
\right )
e^{i (k \theta p_a -2 \pi) \tau(t_a, z)}
\frac{\partial \tau (t_a, z)}{\partial t_a}
\times \nonumber \\
&& \times
\left(
\frac{i (\theta p_{b})^j k \theta u_b}{2 \pi}
+i (\theta u_b)^j
\right)
e^{i (k \theta p_b +2 \pi) \tau(t_b, z)}
\frac{\partial \tau (t_b, z)}{\partial t_b}
\prod_{c \ne a,b} k \theta u_c
e^{i k \theta p_c \tau(t_c, z)}
\frac{\partial \tau (t_c, z)}{\partial t_c}+
\nonumber \\
&& + \sum_{a=1}^n
\frac{i (\theta p_{a})^i}{2 \pi}
\frac{i (\theta p_{a})^j}{2 \pi}
k \theta u_a
e^{i k \theta p_a \tau(t_a, z)}
\frac{\partial \tau (t_a, z)}{\partial t_a}
\prod_{c \ne a} k \theta u_c
e^{i k \theta p_c \tau(t_c, z)}
\frac{\partial \tau (t_c, z)}{\partial t_c}
\Bigg].
\label{T^ij-1}
\end{eqnarray}
The factor $i (\theta p_{a})^i k \theta u_a$ corresponds to
$\theta^{im} \partial_m A_n l^n$ in the coordinate basis.
We will show how this becomes the covariant form
$\theta^{im} F_{mn} l^n$. Note that
\begin{equation}
\frac{i (\theta p)^i k \theta u}{2 \pi}
\mp i (\theta u)^i
= \frac{i (\theta p)^i k \theta u -i (\theta u)^i (k \theta p)}
{2 \pi}
+ \frac{(\theta u)^i}{2 \pi} i ( k \theta p \mp 2 \pi ),
\end{equation}
where the first term on the right-hand side corresponds to
$\theta^{im} (\partial_m A_n -\partial_n A_m )l^n$
divided by $2 \pi$
in the coordinate basis and the second term is a total derivative:
\begin{equation}
\frac{(\theta u)^i}{2 \pi} i ( k \theta p \mp 2 \pi )
e^{i (k \theta p \mp 2 \pi) \tau}
= \frac{(\theta u)^i}{2 \pi}
\frac{\partial}{\partial \tau}
e^{i (k \theta p \mp 2 \pi) \tau}.
\end{equation}
Now consider surface terms in the path-ordered exponential
when we integrate with respect to $\tau_n$. We have
\begin{eqnarray}
&& \int_{\tau_{n-2}}^1 d\tau_{n-1}~
i l^n A_n (x+l \tau_{n-1})
\int_{\tau_{n-1}}^1 d\tau_{n}
\frac{\partial}{\partial \tau_n} \left(
\frac{i \theta^{im}}{2 \pi} A_m (x+l \tau_n)
e^{\mp 2 \pi i \tau_n}
\right)
\times \nonumber \\
&& \times \int_{\tau_{n}}^1 d\tau_{n+1}~
i l^n A_n (x+l \tau_{n+1})
\int_{\tau_{n+1}}^1 d\tau_{n+2} \ldots
\nonumber \\
&=& - \int_{\tau_{n-2}}^1 d\tau_{n-1}~
\frac{i^2 \theta^{im}}{2 \pi} l^n A_n (x+l \tau_{n-1})
A_m (x+l \tau_{n-1}) e^{\mp 2 \pi i \tau_{n-1}}
\times \nonumber \\
&& \times \int_{\tau_{n-1}}^1 d\tau_{n+1}~
i l^n A_n (x+l \tau_{n+1})
\int_{\tau_{n+1}}^1 d\tau_{n+2} \ldots
\nonumber \\
&& + \int_{\tau_{n-2}}^1 d\tau_{n-1}~
i l^n A_n (x+l \tau_{n-1})
\times \nonumber \\
&& \times 
\int_{\tau_{n-1}}^1 d\tau_{n}~
\frac{i^2 \theta^{im}}{2 \pi} A_m (x+l \tau_{n})
l^n A_n (x+l \tau_{n}) e^{\mp 2 \pi i \tau_n}
\int_{\tau_{n}}^1 d\tau_{n+2} \ldots,
\end{eqnarray}
where the $\ast$ product is implicit.
The surface contributions precisely provide the nonlinear
terms $i A_m \ast A_n - i A_n \ast A_m$ in $F_{mn}$.
Thus we have shown that the first part in (\ref{T^ij-1})
can be written in the coordinate basis as follows:
\begin{equation}
\frac{i \theta^{im}}{2 \pi}
\frac{i \theta^{jn}}{2 \pi}
\int_0^1 d\tau'~ e^{-2 \pi i \tau'} l^{m'} F_{mm'} (x+l \tau')
\int_0^1 d\tau''~ e^{2 \pi i \tau''} l^{n'} F_{nn'} (x+l \tau'').
\label{first-term-T^ij}
\end{equation}
The second part in (\ref{T^ij-1}) contains the factor
which corresponds to
$\theta^{im} \theta^{jn} l^{n'} \partial_m \partial_n A_{n'}$
in the coordinate basis.
We expect that this factor turns into
the gauge covariant form $\theta^{im}\theta^{jn} l^{n'} D_m
F_{nn'}$ by a manipulation similar to the one described 
in the above and by taking into account
extra surface contributions which arise when
the two integrals with respect to $\tau'$ and $\tau''$
in (\ref{first-term-T^ij}) are next to each other
in the path-ordered exponential.
The result for $A_i$ is
\begin{eqnarray}
&& (z-\bar{z})^2 \langle
\partial X^i \bar{\partial} X^j e^{ikX(z)}
\exp \left(
i \int_{-\infty}^{\infty} dt~ A_i (X) \frac{dX^i}{dt}
\right) \rangle \nonumber \\
&=& \int dx \ast \Bigg[ e^{ikx}
\exp \left(
i \int_0^1 d\tau~ l^i A_i (x+l \tau) \right)
\times \nonumber \\
&& \times
\Bigg\{ \frac{i \theta^{im}}{2 \pi}
\frac{i \theta^{jn}}{2 \pi}
\int_0^1 d\tau'~ e^{-2 \pi i \tau'} l^{m'} F_{mm'} (x+l \tau')
\int_0^1 d\tau''~ e^{2 \pi i \tau''} l^{n'} F_{nn'} (x+l \tau'')
+ \nonumber \\
&& +
\frac{i \theta^{im} \theta^{jn}}{(2 \pi)^2}
\int_0^1 d\tau'~  l^{n'} D_m F_{nn'} (x+l \tau')
\Bigg\} \Bigg].
\end{eqnarray}

The generalization to the case involving other fields $A_\mu$
and $\Phi^\alpha$ is rather straightforward.
It is obvious that terms involving $A_\mu$ are subleading
in $\alpha'$ just as in the case of the coupling
to the closed string tachyon.
On the other hand, the vertex operator $\partial_\perp X^\alpha$
can produce an $O(1)$ contribution if it is contracted with
the $e^{ikX(z)}$ part of the closed string vertex operator.
Therefore, only the first type of the contractions
described at the beginning of this appendix is allowed
for $\partial_\perp X^\alpha$ and the effect of including
the scalar field is
the following replacements in (\ref{T^ij-1})
\begin{eqnarray}
\frac{i (\theta p_{a})^i k \theta u_a}{2 \pi}
\mp i (\theta u_a)^i
&\to&
\frac{i (\theta p_{a})^i k \theta u_a}{2 \pi}
\mp i (\theta u_a)^i
+ \frac{i (\theta p_{a'})^i y_\alpha \phi^\alpha (p_{a'})}{2 \pi},
\label{replacement-1} \\
\frac{i (\theta p_{a})^i}{2 \pi}
\frac{i (\theta p_{a})^j}{2 \pi}
k \theta u_a
&\to&
\frac{i (\theta p_{a})^i}{2 \pi}
\frac{i (\theta p_{a})^j}{2 \pi}
k \theta u_a
+ \frac{i (\theta p_{a'})^i}{2 \pi}
\frac{i (\theta p_{a'})^j}{2 \pi}
y_\alpha \phi^\alpha (p_{a'}),
\label{replacement-2}
\end{eqnarray}
where $a$ labels the vertex operators for $A_i$
and $a'$ the ones for $\Phi^\alpha$.
Since the open Wilson line now involves $\Phi^\alpha$
as well as $A_i$ there are additional contributions
to the surface terms of the $\tau$ integration
in the process of making $\theta^{im} \partial_m A_n l^n$
to $\theta^{im} F_{mn} l^n$.
They are precisely necessary terms to make
$\theta^{im} \partial_m \Phi^\alpha y_\alpha$
in the last term of (\ref{replacement-1})
to $\theta^{im} D_m \Phi^\alpha y_\alpha$
and we also expect that
$\theta^{im} \theta^{jn}
\partial_m \partial_n \Phi^\alpha y_\alpha$
in (\ref{replacement-2}) becomes
$\theta^{im} \theta^{jn}
D_m D_n \Phi^\alpha y_\alpha$.
After symmetrizing the indices $i$ and $j$,
we obtain the expression for $T^{ij}$
(\ref{emforkneq}).

\section{Conservation of Energy-Momentum Tensor}
We prove
\begin{equation}
k_i T^{ij} + k_\alpha T^{\alpha j} =0,
\label{conservation}
\end{equation}
in bosonic string theory.
Let us begin with some preparations.
It is convenient to define
\begin{equation}
D_\tau {\cal O} (x+l \tau)
\equiv l^i D_i {\cal O} (x+l \tau)
+i [ y_\alpha \Phi^\alpha, {\cal O}] (x+l \tau).
\end{equation}
Since
\begin{eqnarray}
&& \ast \Bigg[
\exp \left( i \int_0^1 d\tau~ l^i A_i (x + l \tau)
+ y_\alpha \Phi^\alpha (x + l \tau) \right)
\int_0^1 d\tau'~ D_{\tau'} {\cal O} (x+l \tau') \Bigg]
\nonumber \\
&=& \ast \Bigg[
\exp \left( i \int_0^1 d\tau~ l^i A_i (x + l \tau)
+ y_\alpha \Phi^\alpha (x + l \tau) \right)
{\cal O} (x+l)
\nonumber \\
&& \qquad - {\cal O} (x)
\exp \left( i \int_0^1 d\tau~ l^i A_i (x + l \tau)
+ y_\alpha \Phi^\alpha (x + l \tau) \right) \Bigg],
\end{eqnarray}
where the path ordering is implicit, we have
\begin{equation}
\int dx~ \ast \Bigg[
e^{ikx} \exp \left( i \int_0^1 d\tau~ l^i A_i (x + l \tau)
+ y_\alpha \Phi^\alpha (x + l \tau) \right)
\int_0^1 d\tau'~ D_{\tau'} {\cal O} (x+l \tau') \Bigg] =0.
\end{equation}
The following formula is also useful:
\begin{eqnarray}
&& \ast \Bigg[
\exp \left( i \int_0^1 d\tau~ l^i A_i (x + l \tau)
+ y_\alpha \Phi^\alpha (x + l \tau) \right)
\times \nonumber \\
&& \times
\int_0^1 d\tau'~ D_{\tau'} f (x+l \tau')
\int_0^1 d\tau''~ g (x+l \tau'')
\Bigg]
\nonumber \\
&=& \ast \Bigg[
\exp \left( i \int_0^1 d\tau~ l^i A_i (x + l \tau)
+ y_\alpha \Phi^\alpha (x + l \tau) \right)
\times \nonumber \\
&& \times \left\{
\int_0^1 d\tau' [f,g] (x+l \tau')
+\int_0^1 d\tau' g (x+l \tau') f (x+l)
- f(x) \int_0^1 d\tau' g (x+l \tau') \right\}
\Bigg]. \nonumber \\
\label{commutator-formula}
\end{eqnarray}
Now consider
\begin{eqnarray}
k_i T^{ij}
&=& \frac{\theta^{jn}}{(2 \pi)^2}
\int dx~ \ast \Bigg[ e^{ikx}
\exp \left( i \int_0^1 d\tau~ l^i A_i (x + l \tau)
+ y_\alpha \Phi^\alpha (x + l \tau) \right)
\times \nonumber \\
&& \times \Bigg\{
i \int_0^1 d \tau'~ e^{-2 \pi i \tau'}
y_\alpha D_{\tau'} \Phi^\alpha (x + l \tau')
\times \nonumber \\
&& \times
i \int_0^1 d \tau''~ e^{2 \pi i \tau''} \left(
l^{n'} F_{nn'} (x+ l \tau'')
+ y_\beta D_n \Phi^\beta (x + l \tau'') \right)
+ \nonumber \\
&& +
i \int_0^1 d \tau'~ e^{-2 \pi i \tau'} \left(
l^{m'} F_{nm'} (x+ l \tau')
+ y_\alpha D_n \Phi^\alpha (x + l \tau') \right)
\times \nonumber \\
&& \times
i \int_0^1 d \tau''~ e^{2 \pi i \tau''}
y_\beta D_{\tau''} \Phi^\beta (x + l \tau'')
+ \nonumber \\
&& +
i \int_0^1 d \tau'~ \Big(
l^{n'} l^m D_m F_{nn'} (x+ l \tau')
\nonumber \\ &&
+ y_\alpha l^m D_m D_n \Phi^\alpha (x + l \tau') 
+ y_\alpha l^m D_n D_m \Phi^\alpha (x + l \tau') 
\Big) \Bigg\} \Bigg],
\label{kT}
\end{eqnarray}
where we have used that
\begin{equation}
y_\alpha l^m D_m \Phi^\alpha (x + l \tau )
= y_\alpha D_\tau \Phi^\alpha (x + l \tau ).
\end{equation}
There are two different contributions when we make
partial integrations with respect to $D_{\tau'}$
in the second line and $D_{\tau''}$
in the fifth line.
The first one is the one which arises when
$D_{\tau'}$ ($D_{\tau''}$) acts on
$e^{-2 \pi i \tau'}$ ($e^{2 \pi i \tau''}$)
and this is precisely canceled by $k_\alpha T^{\alpha j}$.
The other one can be evaluated using the formula
(\ref{commutator-formula}) and gives a factor
\begin{equation}
2 i^2 \int_0^1 d \tau' [ y_\alpha \Phi^\alpha,
l^{n'} F_{nn'} + y_\beta D_n \Phi^\beta].
\label{from-commutator}
\end{equation}
The three terms in the last two lines of (\ref{kT})
can be written as
\begin{eqnarray}
l^{n'} l^m D_m F_{nn'} (x+ l \tau')
&=& l^{n'} D_{\tau'} F_{nn'} (x+ l \tau')
-i [y_\alpha \Phi^\alpha, l^{n'} F_{nn'}] (x + l \tau'),
\nonumber \\
y_\alpha l^m D_m D_n \Phi^\alpha (x + l \tau')
&=& y_\alpha D_{\tau'} D_n \Phi^\alpha (x + l \tau')
-i [y_\beta \Phi^\beta, y_\alpha D_n \Phi^\alpha] (x + l \tau'),
\nonumber \\
y_\alpha l^m D_n D_m \Phi^\alpha (x + l \tau')
&=& y_\alpha l^m D_m D_n \Phi^\alpha (x + l \tau')
+i [l^m F_{nm}, y_\alpha \Phi^\alpha] (x + l \tau'),
\nonumber \\
&&
\end{eqnarray}
so that the last integral of (\ref{kT}) gives
\begin{equation}
i \int_0^1 d \tau' \left(
-2i [ y_\alpha \Phi^\alpha, l^{n'} F_{nn'}]
-2i [y_\beta \Phi^\beta, y_\alpha D_n \Phi^\alpha]
\right),
\end{equation}
which cancels (\ref{from-commutator}).
Thus we have shown the conservation (\ref{conservation}).

\newpage
\renewcommand{\baselinestretch}{0.87}

\begingroup\raggedright\endgroup
\end{document}